\documentclass[12pt]{iopart}

\usepackage{graphicx}

\begin{document}

\title[]{Scale invariant quantum potential leading to globally self-trapped wave function in Madelung fluid}

\author{Agung Budiyono}

\address{Institute for the Physical and Chemical Research, RIKEN, 2-1 Hirosawa, Wako-shi, Saitama 351-0198, Japan}
\ead{agungby@riken.jp}

\begin{abstract}
We show in spatially one dimensional Madelung fluid that a simple requirement on local stability of the maximum of quantum probability density will, if combined with the global scale invariance of quantum potential, lead to a class of quantum probability densities globally being self-trapped by their own self-generated quantum potentials, possessing only a finite-size spatial support. It turns out to belong to a class of the most probable wave function given its energy through the maximum entropy principle.  We proceed to show that there is a limiting case in which the quantum probability density becomes the stationary-moving soliton-like solution of the Schr\"odinger equation.  
\end{abstract}

\pacs{03.65.Ge,03.65.Ca,03.65.Vf}
\maketitle

\section{Madelung fluid: global scale invariant quantum potential}

Let us consider a spatially one dimensional Madelung fluid \cite{Madelung paper}  for a single free particle with mass $m$. The state of the system is then determined by a pair of fields in position space $q$ as:  $\{\rho(q),v(q)\}$, where $\rho(q)$ is a normalized quantity called as quantum probability density and $v(q)$ is velocity field. The temporal evolution of both at time $t$ is then assumed to satisfy the following coupled dynamical equation:
\begin{eqnarray}
m\frac{dv}{dt}=-\partial_qU,\hspace{3mm}\partial_t\rho+\partial_q(\rho v)=0.
\label{Madelung fluid dynamics}
\end{eqnarray}
Here, $U(q)$ is the so-called quantum potential generated by the quantum amplitude $R=\rho^{1/2}$ as 
\begin{equation}
U(q)=-\frac{\hbar^2}{2m}\frac{\partial_q^2R}{R},
\label{quantum potential}
\end{equation}  

For the case of a spatially one dimensional fluid, the velocity field can always be written as the spatial gradient of a scalar function $S(q)$ as 
\begin{equation}
v(q)=\partial_qS/m. 
\label{velocity field vs quantum phase}
\end{equation}
One can then use this new quantity to define a complex-valued wave function $\psi=R\exp(iS/\hbar)$ to show that the Madelung fluid dynamics given in Eq. (\ref{Madelung fluid dynamics}) is equivalent to the Schr\"odinger equation for a single free particle as follows:
\begin{equation}
i\hbar\partial_t\psi(q;t)=-\frac{\hbar^2}{2m}\partial_q^2\psi(q;t).
\label{Schroedinger equation}
\end{equation}

Notice that while the equation on the right of Eq. (\ref{Madelung fluid dynamics}) is but the conventional continuity equation which guarantees the conservation of probability flow,  the left equation takes the form of Newtonian dynamical equation with a classically absence new term appears on the right hand side. In this regards, the term $F=-\partial_qU$ is called as quantum force. This fact suggests that the quantum force is  the new quantity which is responsible for the nonclassical behaviors of Schr\"odinger equation. Thus, it is  reasonable to pay  serious attention to the the property of the quantum potential. 

Let us mention an important properties of quantum potential that will play important roles in our discussion later. First is multiplying the quantum probability density with a constant will not change the profile of the quantum potential. Namely, the quantum potential is invariant under global rescaling of the quantum probability density, namely its own source, as: 
\begin{equation}
U(c\rho(q))=U(\rho(q)),
\end{equation}
where $c$ is constant. This shows that the quantum potential only cares about the form of the quantum probability density and is independent from the strength of the  latter  \cite{Bohm-Hiley book}. It is as if the quantum potential considers the quantum probability density as  a code in telecommunication system in case of which only the profile of the sequence of the binary wave is important, the strength of the received wave itself is of no use.  In this sense, the quantum potential is of informational nature. For an interesting and stimulating discussion concerning this matter see \cite{Bohm-Hiley book}.

Among the consequences of the above invariant property is that, first rescaling the quantum probability density by the mass of the particle, $\tilde{\rho}=m\rho$ will not change the dynamical equation on the left of Eq. (\ref{Madelung fluid dynamics}). On the other hand, the continuity equation on the right of Eq. (\ref{Madelung fluid dynamics}) becomes
\begin{equation}
\partial_t\tilde{\rho}+\partial_q({v\tilde{\rho}})=0,
\label{continuity equation for mass density}
\end{equation}
which can now be read as the equation for the conservation of mass density, rather than the conservation of quantum probability density.  The other consequence of the scale invariance property of quantum potential is that even at points where the strength of the quantum probability density is very low, the quantum potential that it generates at that point might be very high. In this paper, we shall be interested in a class of quantum probability densities with this specific property.

Next, let us mention another property of quantum potential $U(q)$ that the average of the quantum force, $F=-\partial_qU$, over the quantum probability is vanishing \cite{Holland book}
\begin{equation}
\int dq\hspace{1mm}\partial_qU(q)\rho(q)=0.
\label{vanishing average imaginability}
\end{equation}
This can be proven easily by assuming $\rho(\pm\infty)=0$. Imposing this into the dynamical equation of Eq. (\ref{Madelung fluid dynamics}), one reproduces the Ehrenfest theorem \cite{Holland book}
\begin{equation}
m\frac{d{\bar v}}{dt}=0,
\label{Ehrenfest theorem}
\end{equation}
where ${\bar v}$ is the average value of the velocity field defined as ${\bar v}\equiv\int dq\hspace{1mm}v(q)\rho(q)$. 

\section{Local stability and globally self-trapping quantum potential}

Let us now show that local geometrical restriction on the maximum point of the quantum probability density, if combined with the scale invariance of the quantum potential will determine the global geometrical property of the quantum potential, thus the quantum probability density as well. First, since the quantum probability density is vanishing at infinity, $\rho(\pm \infty)=0$, non-negative and normalized, it must at least have one local maximum point. Let us denote this maximum point by $q=Q$. It thus satisfies 
\begin{equation}
\partial_q\rho|_{Q}=0,\hspace{2mm}\partial_{q}^2\rho|_{Q}<0.
\label{principle of maximal imagination} 
\end{equation}
One first observes that at this point the quantum potential is positive definite
\begin{equation}
U(Q)=-\frac{\hbar^2}{4m}\Big(-\frac{1}{2}\Big(\frac{\partial_q\rho}{\rho}\Big)^2+\frac{\partial_q^2\rho}{\rho}\Big)\Big|_Q>0.
\end{equation} 
Before proceeding, let us write a useful formula for later discussion 
\begin{equation}
\frac{\partial_q^n\rho^s}{\rho^s}\Big|_{Q}=s\frac{\partial_q^n\rho}{\rho}\Big|_{Q},
\label{identity for realism}
\end{equation}
which can be shown easily by utilizing the left equation in (\ref{principle of maximal imagination}) to be valid for any positive integer $n$. 

Now, let us put a local restriction on a class of quantum probability densities $\rho(q)$ so that its maximum point stays at the minimum point of the quantum potential $U(q)$ which it generates through Eq. (\ref{quantum potential}). One therefore imposes  
\begin{equation}
\partial_q U|_{Q}=0,\hspace{5mm}\partial_{q}^2U|_{Q}\ge 0.
\label{stable stationary stochastic potential}
\end{equation}   
Dynamically we are thus looking for a class of quantum probability densities in which at least its maximum is temporally stable.  Next, let us show that the restrictions given by Eqs. (\ref{stable stationary stochastic potential}) will uniquely determine the form of $U(q)$ as a function of $\rho(q)$. First, from the global scale invariant property of the quantum potential, then the quantum force is also global scaling invariance; so that one has $\partial_qU(c\rho)=\partial_qU(\rho)$ for any real constant $c$.  It is therefore reasonable to write the quantum force to take the following non-trivial form:  
\begin{equation}
\partial_qU(\rho)=\frac{1}{\rho^s}(a_0+a_1\partial_q+a_2\partial_q^2+a_3\partial_q^3+\dots)\rho^s,
\label{stationary state 1}
\end{equation}
where $s$ and $a_i$, $i=0,1,2,\dots,$ are arbitrary real number. Evaluating at $q=Q$ and using the fact of Eq. (\ref{identity for realism}) one has 
\begin{equation}
\partial_qU(\rho(Q))=a_0+\frac{s}{\rho}(a_1\partial_q+a_2\partial_q^2+a_3\partial_q^3+\dots)\rho\Big|_{Q}.
\label{stationary state 2}
\end{equation}
The left equation in (\ref{stable stationary stochastic potential}) imposes the right hand side of Eq. (\ref{stationary state 2}) to be vanishing. Keeping in mind Eqs. (\ref{principle of maximal imagination}) and the fact that $\partial_q^n\rho|_{Q}$, for  $n\ge 3$, are fluctuating between positive and negative value, $\partial_qU|_{Q}=0$ can then be accomplished by imposing $a_0=0$, $a_j=0$ for $j\ge 2$, and $a_1$ is arbitrary, yet non-vanishing. One therefore has
\begin{equation}
\partial_qU(\rho)=a_1\frac{\partial_q\rho^s}{\rho^s}.
\label{stationary state 3}
\end{equation}

Next, let us rewrite Eq. (\ref{stationary state 3}) as follows
\begin{equation}
\partial_qU(\rho)=a_1\partial_q\ln\rho^s=a_1s\partial_q\ln\rho=a_1s\frac{\partial_q\rho}{\rho}=a_s\frac{\partial_q\rho}{\rho},
\label{stationary state 4}
\end{equation}
where we have denoted $a_s=a_1s$. Now, taking spatial derivation on both sides of the above equation and using the left equation in Eq. (\ref{principle of maximal imagination}), one gets 
\begin{equation}
\partial_q^2U\Big|_{Q}=a_s\frac{\partial_q^2\rho}{\rho}\Big|_{Q}-a_s\Big(\frac{\partial_q\rho}{\rho}\Big)^2\Big|_{Q}=a_s\frac{\partial_q^2\rho}{\rho}\Big|_{Q}.
\end{equation}
Comparing this fact to the right inequality in (\ref{stable stationary  stochastic potential}) and keeping in mind the fact that $\partial_q^2\rho|_{Q}<0$, one concludes that $a_s$ must be non-positive. One can then verify that any quantum probability density that satisfies Eq. (\ref{stationary state 4}) satisfies all the requirements that we set at the beginning. Moreover, assuming $\rho(\pm\infty)=0$, Ehrenfest theorem is automatically satisfied
\begin{equation}
\int dq\hspace{1mm}\rho\partial_qU=a_s\int dq\hspace{1mm}\partial_q\rho=0.
\end{equation}

To proceed, for simplicity of notation, let us rewrite Eq. (\ref{stationary state 4}) as follows
\begin{equation}
\partial_qU=-a\frac{\partial_q\rho}{\rho},\hspace{2mm}a\ge 0.
\label{scale invariance - self-trapped state}
\end{equation}
It can be readily integrated to obtain
\begin{equation}
\rho(q;a)=\frac{1}{Z(a)}\exp\Big(-\frac{1}{a}U(q;a)\Big),
\label{MBG canonical distribution}
\end{equation}
where $Z(a)=\int dq\hspace{1mm}\exp(-U/a)$  is a normalization constant  independent of $q$.  We shall show later that $U(q)$ can be interpreted as internal energy density. Bearing this in mind, then the quantum probability density given in Eq. (\ref{MBG canonical distribution})  resembles in form  with the Maxwell-Boltzmann-Gibbs  (MBG)   canonical  distribution in equilibrium thermodynamics.  It is thus  suggestive to apply thermodynamics formalism to further study the property of quantum probability density given in Eq. (\ref{MBG canonical distribution}) \cite{AgungPRE}. 

Next, let us recall that in quantum mechanics $\rho(q)$ gives the essential information on the position of the particle \cite{Bohm-Hiley book,Isham book,Bell unspeakable}.  In the so-called pragmatical approach of quantum mechanics, $\rho(q)$ is given meaning as the probability density that the particle will be found at $q$ if a measurement is performed. On the other hand, in the ontological approach, $\rho(q)$ is argued as the probability density that the particle is at $q$ regardless of any measurement. It is thus reasonable to quantify the randomness encoded in $\rho(q)$. One obvious way is then to use the differential entropy or the so-called Shannon information entropy \cite{Shannon entropy} over the quantum probability density given by 
\begin{equation}
H[\rho]=-\int dq\hspace{1mm}\rho(q) \ln \rho(q). 
\end{equation}
It gives the degree of localization of the wave function in position space. 

One can then show that the canonical quantum probability density of the form given in Eq. (\ref{MBG canonical distribution}) maximizes the Shannon entropy provided that the average quantum potential is given by \cite{Mackey-MEP}
\begin{equation}
{\bar U}=\int dq\hspace{1mm}U(q)\rho(q).
\label{average quantum potential}
\end{equation}
Hence, the quantum probability density developed in the previous section satisfies the so-called maximum entropy principle \cite{Jaynes-MEP}. It has been argued that the maximum entropy principle is the only method to infer from an incomplete information, which does not lead to logical inconsistency \cite{Shore-Johnson-MEP}.  The self-trapped quantum probability density can then be seen as the most probable quantum probability density given its average quantum and kinetic energy \cite{AgungPRA1}. 
 
\begin{figure}[htbp] 
   \centering
  \includegraphics[width=7cm]{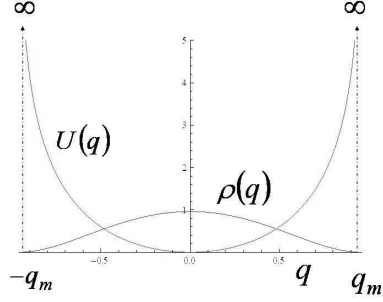} 
   \caption{The profile of quantum probability density and its corresponding quantum potential which satisfies Eq. (\ref{NPDE for U}).}
   \label{self-trapped QPD}
\end{figure}

Combined with the definition of quantum potential given in Eq. (\ref{quantum potential}), Eq. (\ref{MBG canonical distribution}) comprises a differential equation for $\rho(q)$ or $U(q)$ subjected to the condition that $\rho(q)$ must be normalized, $\int dq\rho(q)=1$. In term of quantum potential, one has the following nonlinear differential equation
\begin{equation}
\partial_q^2U-\frac{1}{2a}(\partial_qU)^2-\frac{4ma}{\hbar^2}U=0.
\label{NPDE for U}
\end{equation}
Figure \ref{self-trapped QPD} shows the solution of Eq. (\ref{NPDE for U}) with the boundary conditions: $\partial_qU(0)=0$ and $U(0)=1$ for $a=1$.  All numerical solutions in this  paper are obtained by putting $m=\hbar=1$.  The quantum potential is shifted down so that its global minimum is vanishing. One can first see that the maximum point of the quantum probability density and the minimum point of the corresponding quantum potential coincide, thus satisfies our requirement. Yet, what makes even interesting is that, though we only requires the quantum potential to trap the area of the quantum probability density around its maximum, it turns out that the resulting  quantum potential is convex everywhere and becomes the global trapping potential for its own source: quantum probability density.  

\begin{figure}[tbp]
\begin{center}
\includegraphics*[width=9cm]{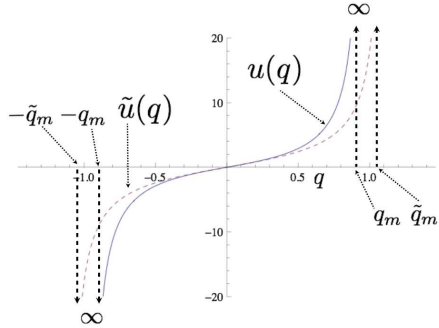}
\end{center}
\caption{$\tilde{u}(q)$ and $u(q)$. See text for detail.}
\label{tangent function vs blowing-up}
\end{figure}

Next, one can also see that the solution plotted in Fig. \ref{self-trapped QPD} possesses blowing-up points at $q=\pm q_m$,  namely $U(\pm q_m)=\infty$ \cite{blowing-up NDE}. Let us first prove that the blowing-up will certainly occur at finite point from the origin as along as $U(0)\equiv X$ is not vanishing. To do this, Let us define a new variable $u=\partial_qU$. The nonlinear differential equation of Eq. (\ref{NPDE for U}) then transforms into 
\begin{equation}
\partial_qu=\frac{1}{2a}u^2+\frac{4ma}{\hbar^2}U,
\label{blowing-up NPDE for u}
\end{equation}
Moreover, the boundary condition translates into $u(0)=\partial_qU(0)=0$. Let us now consider the following nonlinear differential equation
\begin{equation}
\partial_q\tilde{u}=\frac{1}{2a}\tilde{u}^2+\frac{4ma}{\hbar^2}X, 
\label{blowing-up for inferior solution}
\end{equation}
where $X\equiv U(0)$ with $\tilde{u}(0)=0$. Since $U(q)\ge U(0)=X$, then it is obvious that $|u(q)|\ge|\tilde{u}(q)|$. 

On the other hand, one can solve the latter differential equation of Eq. (\ref{blowing-up for inferior solution}) analytically to have
\begin{equation}
\tilde{u}(q)=d\tan(gq), \hspace{2mm}d=\frac{2a}{\hbar}\sqrt{2m}, \hspace{2mm}g=\frac{1}{\hbar}\sqrt{2mX}.  
\label{inferior solution}
\end{equation}
It is then clear that at $q=\pm\tilde{q}_m=\pm\pi/(2g)$, $\tilde{u}$ is blowing-up, namely $\tilde{u}(\pm\tilde{q}_m)=\pm\infty$. Recalling the fact that $|u(q)|\ge |\tilde{u}(q)|$, then $u(q)$ is also blowing-up at point $q=\pm q_m$, $u(\pm q_m)=\pm\infty$, where $q_m\le \tilde{q}_m$. See Fig. \ref{tangent function vs blowing-up}. Putting this into the original nonlinear differential equation of Eq. (\ref{NPDE for U}), one concludes that $U(q)$ is also blowing up at $q=\pm q_m$, $U(\pm q_m)=\infty$. Notice that even though $u(-q_m)=-\infty$ is blowing-up to minus infinity, $U(-q_m)=\infty$ is obviously blowing up into positive infinity. Next, it is clear that the blowing-up is due to the existence of the nonlinear term on the right hand side of Eq. (\ref{NPDE for U}). Hence, finally one can safely say that for any non-vanishing $U(0)=X$, the corresponding quantum probability density possesses only a finite range of support, $q\in [-q_m,q_m]$. The case when $U(0)=0$ will give the trivial solution $U(q)=0$ for the whole space $q$ so that $\rho(q)$ is unnormalizable. The above fact also confirms our assertion in Section I that the quantum potential might take large value even at points where the corresponding quantum probability density is very small.

\begin{figure}[htbp] 
   \centering
  \includegraphics[width=7cm]{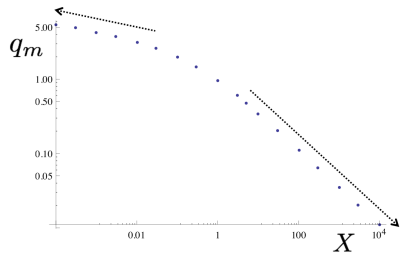} 
   \caption{The half length of support $q_m$ plotted against the variation of the global minimum $U(0)=X$.}
   \label{blowing-up point vs X}
\end{figure}

In Fig. \ref{blowing-up point vs X} we plot the blowing-up point $q=q_m$, namely half length of the spatial support of the quantum probability density against the value of the quantum potential at the global minimum: $U(0)=X$. One observes that $q_m$ is decreasing as we increase $X$ for fixed $a=1$. This can be understood directly from Eq. (\ref{inferior solution}). One can also confirm that the occurrence of blowing-up is the case only when $X\neq 0$, namely $\lim_{X\rightarrow 0}q_m=\infty$. 

\section{Solitonic wave function} 

\begin{figure}[htbp]   
 \centering
  \includegraphics[width=7cm]{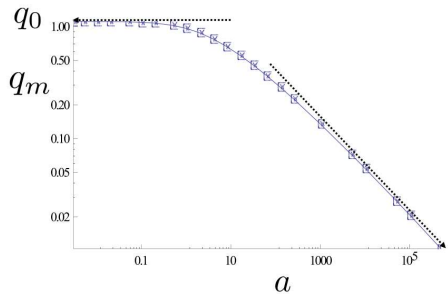} 
   \caption{The half length of the support against the variation of $a$.}
   \label{a vs half length of support}
\end{figure}

Now let us proceed to study the behavior of the quantum potential as one varies the non-negative parameter $a$. Figure \ref{a vs half length of support} gives the variation of the blowing-up point, $q_m$, thus the size of the range of the support against the variation of the parameter $a$. This is obtained by solving the differential equation of Eq. (\ref{NPDE for U}) with fixed boundary conditions: $\partial_qU(0)=0$ and $U(0)=1$. One first observes that as $a$ is increased, $q_m$ decreases and eventually vanishing for infinite value of $a$. This shows that the quantum probability density is becoming narrower for larger $a$ while kept normalized; and eventually collapsing onto Dirac delta function for infinite value of $a$. 

A very interesting phenomena is observed as one decreases the parameter $a$ toward zero. One finds that the blowing up point $q_m$ is increasing and eventually converging toward a finite value $q_0$ for $a=0$, 
\begin{equation}
\lim_{a\rightarrow 0}q_m(a)=q_0.
\label{finite blowing-up point at vanishing temperature}
\end{equation}
This suggests to us that the quantum potential and the corresponding quantum probability density are also converging toward certain functions for vanishing value of $a$:
\begin{equation}
\lim_{a\rightarrow 0}U(q;a)=U_0(q),\hspace{3mm}
\lim_{a\rightarrow 0}\rho(q;a)=\rho_0(q). 
\label{vanishing temperature QP and QPD}
\end{equation}

\begin{figure}[htbp]   
 \centering
  \includegraphics[width=7cm]{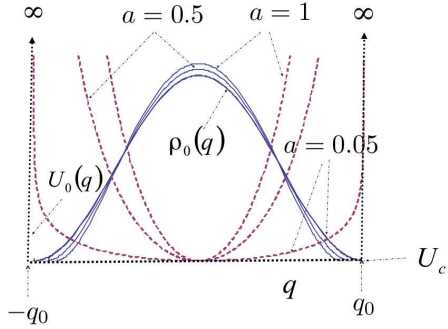} 
   \caption{The profile of self-trapped quantum probability density and its corresponding quantum potential for several small values of parameter $a$. We also plot the case when $a=0$ obtained analytically in Eq. (\ref{localized-stationary QPD}).}
   \label{small temperature QPD and QP}
\end{figure}

Let us discuss this situation in more detail.  In Figure \ref{small temperature QPD and QP} we plot the profile of the quantum probability density and the corresponding quantum potential for several small values of parameter $a$ with fixed boundary conditions: $\partial_qU(0)=0$ and $U(0)=1$. One can then see that as $a$ is decreased, the quantum potential is becoming flatterer inside the support before blowing-up at $q=\pm q_m(a)$. One might then guess that at the limit $a=0$, the quantum potential is perfectly flat inside the support and is infinite at the blowing-up points, $q=\pm q_0$. Let us show that this guess is correct. To do this, let us denote the assumed constant value of the quantum potential inside the support as $U_c$. Recalling the definition of quantum potential given in Eq. (\ref{quantum potential}), one has 
\begin{equation}
-\frac{\hbar^2}{2m}R_0(q)=U_cR_0(q), 
\label{stationary Schroedinger equation 1}
\end{equation}
where $R_0\equiv \rho_{0}^{1/2}$.  This has to be subjected to the condition that $R_0(\pm q_0)=0$.  

Next, solving equation (\ref{stationary Schroedinger equation 1}) one obtains
\begin{equation}
R_0(q)=A_0\cos(k_0 q),
\label{localized-stationary QPD}
\end{equation}
where $A_0$ is a normalization constant and $k_0$ is related to the quantum potential as 
\begin{equation}
k_0=\sqrt{2mU_c/\hbar^2}. 
\label{stationary wave number}
\end{equation}
The boundary condition imposes $k_0q_0=\pi/2$. In Fig. \ref{small temperature QPD and QP}, we plot the above obtained quantum probability density, $\rho_0(q)$. One can see that as $a$ is decreasing toward zero, $\rho(q;a)$ obtained by solving the differential equation of (\ref{NPDE for U}) is indeed converging toward $\rho_0(q)$ given in equation (\ref{localized-stationary QPD}). This confirms our  guess that at $a=0$ the quantum potential takes a form of flat box with infinite wall at $q=\pm q_0$. 

Let us now take $\{\rho_0(q),v_0(q)\}$ as the initial state of the dynamics. Here $v_0(q;0)=v_c$ is a uniform velocity field with non-vanishing constant value only inside the support. Since at $a=0$ the quantum potential is flat, then inside the support the quantum force is vanishing: $F=-\partial_qU=0$. Inserting this into the dynamical equation of Eqs. (\ref{Madelung fluid dynamics}), one has $dv/dt=0$. Hence the velocity field at infinitesimal lapse of time, $t=\Delta t$, remains constant and uniform. This in turn will not change the initial probability density, but shift it in space by $\Delta q=v_c\Delta t$: $\rho(q;\Delta t)=\rho_0(q-v_c\Delta t;0)$. Accordingly, the support is also shifted by the same amount. This will repeat in the next infinitesimal time lapse and so on and so forth so that at finite lapse of time $t$, both the initial velocity field and quantum probability density remains unchanged but are shifted by an amount $\Delta q=v_ct$. One thus has 
\begin{equation}
\rho(q;t)=\rho_0(q-v_ct;0)=A_0^2\cos^2(k_0q-\omega_0t),
\label{free particle typical-stationary wave function at time t}
\end{equation} 
where we have put $\omega_0=k_0v_c$. 

Before proceeding, let us give physical meaning to the average quantum potential. To do this, let us calculate the ordinary quantum mechanical energy given by $\langle E\rangle\equiv \int_{-q_0}^{q_0} dq\hspace{1mm}\psi^*(q)\big(-\hbar^2/2m\big)\partial_q^2\psi(q)$ . Writing the wave function in polar form one has 
\begin{eqnarray}
\langle E\rangle=\int_{-q_0}^{q_0} dq\hspace{1mm}\Big(-\frac{\hbar^2}{2m}R\partial_q^2R+\frac{1}{2m}R^2(\partial_qS)^2\nonumber\\
-\frac{i\hbar}{m}R\partial_qR\partial_qS-\frac{i\hbar}{2m}R^2\partial_q^2S \Big).
\label{calculation of quantum mechanical energy}
\end{eqnarray} 
The first term on the right hand side is equal to the average quantum potential, $\bar{U}=\int dqU\rho$. Next, defining kinetic energy density as $K(q)=(m/2)v^2(q)$, the second term is equal to the kinetic energy $\bar{K}=\int dq K\rho$ of the Madelung fluid, which for our stationary state is given by ${\bar K}=(m/2)v_c^2$. Further, for a uniform velocity field, the last term is vanishing, $\partial_q^2S=m\partial_qv_c=0$. Again for a uniform velocity field, since $R(q)$ is an even function and $\partial_qR(q)$ is an odd function then the third term is also vanishing. Hence, in total, the quantum mechanical energy of a self-trapped wave function for $a=0$ moving with a uniform velocity field can be decomposed as 
\begin{equation}
\langle E\rangle=\bar{U}_0+\bar{K}.
\label{energy decomposition}
\end{equation}
One can then conclude that $\bar{U}_0$ must essentially be interpreted as the rest energy of the single particle. Namely it is the energy of the particle when it is not moving so that ${\bar K}=0$. Moreover, since inside the support the quantum potential is flat given by $U_c$, one has $\bar{U}_0=\int_{-q_0}^{q_0}U_0(q)\rho(q)=U_c$. Recalling Eq. (\ref{stationary wave number}), one finally obtains
\begin{equation}
\langle E\rangle=\frac{\hbar^2k_0^2}{2m}+\frac{1}{2}mv_c^2.
\label{energy decomposition 2}
\end{equation}

Let us now give the corresponding complex-valued stationary wave function  $\psi(q)$. One thus needs to calculate the quantum phase $S$ which can be obtained by integrating $\partial_qS=mv_c$ to give: $S(q;t)=mv_cq+\sigma(t)$, where $\sigma(t)$ depends only on time. One therefore has $\psi_{st}(q;t)=A_0\cos\big(k_0(q-v_ct))\exp\Big((i/\hbar)(mv_cq+\sigma(t))\Big)$,  where $q\in\mathcal{M}_t\equiv[v_ct-q_0,v_ct+q_0]$. Inserting this into the Schr\"odinger equation of Eq. (\ref{Schroedinger equation}) and using Eq. (\ref{energy decomposition 2}) one has $\sigma(t)=-\langle E\rangle t$ modulo to some constant. Putting all these back, one finally obtains the following solution: 
\begin{equation}
\psi_{st}(q;t)=A_0\cos\big(k_0(q-v_c t))\exp\Big(\frac{i}{\hbar}(mv_c q-\langle E\rangle t)\Big),
\label{typical-stationary wave function}
\end{equation}
where $q\in\mathcal{M}_t$. 

Equation (\ref{typical-stationary wave function}) is of soliton type. This suggests a direct association of the wave function to a particle by considering the wave function as a {\it physical field}. In this regard, the continuity equation of mass density of Eq. (\ref{continuity equation for mass density}) is becoming relevant. On the other hand, we have also shown in the previous section that the localized-stationary-traveling solution belongs to a class of wave function which maximizes Shannon entropy, which suggests that it is a {\it probabilistic wave field}. Hence, one arrives at one of the old problem of quantum mechanics concerning the physical status of the wave function. 

\section{Conclusion and Interpretation}

By exploiting the scaling invariant property of the quantum potential, we show that the requirement of local stability on the maximum of quantum probability density leads us to a class of quantum probability densities which is globally trapped by its own quantum potential with finite-size spatial support. It turns out that they belong to a class of wave function which maximizes Shannon entropy given the average quantum potential. Further, we show that for a single free particle quantum system, there is an asymptotic limit in which the self-trapped wave function is traveling while keeping its form unchanged. This fact thus suggests to us to associate the localized-traveling quantum probability density as a real particle. In contrast to this, in conventional formalism of quantum mechanics, one usually choose a plane wave to represent a free single particle. 

In our formalism of a single particle as a localized and self-trapped wave function, we showed that a particle possesses an internal energy which is absence if one use a plane wave instead. It is equal to the quantum potential. On the other hand, we showed in the beginning of the paper that the quantum potential is invariant under the global rescaling of its own source, namely the quantum probability density. It depends only on the form of the latter. It is a surprising fact, since, usually energy is an extensive quantity with respect to its source. Hence, the internal energy is of different nature from the ordinary one. Since it only records the profile of its own source, one might conclude that its nature is informational, rather than material. 

An interesting point is left unexplored. On one hand, we showed that the canonical form of quantum probability density given in Eq. (\ref{MBG canonical distribution}) is the consequence of the scale invariant property of the quantum potential. On the other hand, we also showed that it maximizes Shannon information entropy given its average quantum potential. One may then expect that these two facts are related in a nontrivial way. 

\ack

\end{document}